\documentclass[aps,pra,superscriptaddress,amsmath,amssymb,preprintnumbers,floatfix,showpacs,12pt]{revtex4}
\usepackage{amssymb}
\usepackage{epsfig}

\begin{document}
\title{ Linear entropy as  an entanglement measure in  two-fermion systems}
\author{Fabrizio Buscemi }
\email{buscemi.fabrizio@unimore.it}
\author{Paolo Bordone}
\affiliation{S3 Research Center,CNR-INFM,  Via Campi 213/A, I-Modena 41100, Italy}
 \affiliation{ Dipartimento di   Fisica, Universit\`{a} di Modena e Reggio Emilia, Modena, Italy}
\author{Andrea Bertoni }
\affiliation{S3 Research Center,CNR-INFM,  Via Campi 213/A, I-Modena 41100, Italy}
\begin{abstract}
We describe an efficient theoretical criterion, suitable for
indistinguishable particles    to quantify the quantum
correlations of any pure two-fermion state, based on the Slater
rank concept. It represents the natural generalization of the
linear entropy  used  to treat  quantum entanglement in
systems of non-identical particles. Such a criterion is here
applied to  an electron-electron  scattering  in a two-dimensional
system in order to perform a quantitative evaluation of the
entanglement dynamics for various spin configurations and to
compare the linear entropy with alternative approaches. Our
numerical results show the dependence of the entanglement
evolution upon   the initial state of the system and its spin
components. The differences with previous analyses accomplished by
using the von Neumann entropy are discussed. The evaluation  of the
entanglement dynamics in terms of the linear entropy results to be
much less demanding from the computational point of view, not
requiring   the diagonalization of the density matrix.

\end{abstract}
\maketitle

\section{Introduction}
The  entanglement, possibly  the most  remarkable feature of
quantum mechanics,  represents  a fundamental resource for quantum
information processing \cite{Schr,Giuli,Peres},  and  the concept
of
 bipartite and multipartite entanglement is nowadays
well-stated  for   quantum systems  composed of  distinguishable
constituents. For an entangled system it is impossible to factor
its  state in a product of  independent states  describing  its
parts. On the other hand  the notion of  entanglement is more
controversial in systems of identical particles \cite{Sch,Shi, Li,
Zanardi, Fisher, You, Vacca,Levay}. The entanglement of such systems is
investigated at present in many areas of physics, like quantum
optics, quantum charge transport in semiconductor, ultracold gases
\cite{brand,zeng,dow,bus,leon,raa,inn}. Here the difficulties appear
in the definition of a criterion  able to classify and quantify
the entanglement.  They  are mainly due to the exchange symmetry
which requires the antisymmetrization or symmetrization of the
quantum wavefunctions describing fermions or bosons, respectively.

 Different methods  to treat  the quantum
entanglement in  systems  of indistinguishable  particles are
present in literature. In the approach developed by Wiseman and
Vaccaro ~\cite{Vacca}  the entanglement of the particles
is a sort of accessible entanglement, i.e. the maximum value of
the entanglement that could be extracted  from the system and
placed in quantum registers, from which it could be used to
perform quantum information processing.  In the theory  introduced
by Zanardi the  entanglement should be evaluated by using the
density matrix in a mode-occupation represention  and is based on
the formal mapping of the Fock space into states  of qubits
\cite{Zanardi}. The method  proposed by  Schliemann is based on
the Slater rank of the state (i.e. the minimum number of Slater
determinants needed to express it) as a counterpart of the Schmidt
rank criterion usually adopted for  distinguishable particles
\cite{Sch,schli2,Schl3}. This latest approach  has been recently
reexamined by other  authors, which,  for the case of a
two-particle pure state,  have suggested to  evaluate
quantitatively the entanglement as the  von Neumann entropy (vNE)
of the one-particle  reduced density matrix \cite{You,ghira,sand}.
Here the quantum correlations due to symmetrization or
antisymmetrization of the wavefunction do not represent a genuine
manifestation of  quantum entanglement \cite{ghira2}. Therefore
from this point of view such a criterion results to be  the
natural generalization of the approach commonly used to treat
quantum correlations  in  systems  of distinguishable particles.

Following the basic concepts of the approach proposed by
Schliemann \cite{Sch}  in this paper we  discuss  an
entanglement criterion for two-fermion systems. It is still   based on the
analogous of the Schmidt decomposition theorem for the pure
fermion state but  requires the calculation of the linear entropy (LE) of the
one-particle reduced density matrix. We   intend  to establish whether the LE
can be considered   a valid measure of the lack of knowledge about
the quantum state describing the system, which should include not
only  the uncertainty  due to the impossibility of attributing a
definite state to each particle but also the amount of uncertainty
deriving from the indistinguishability of the particles. In order
to get a  better understanding of  this criterion,  we first apply
it to study
 a simple prototype theoretical model, then  we analyze a system of physical interest,
namely a two-electron scattering event in a  2D semiconductor
structure. The  entanglement dynamics  in such a system has been
recently investigated in terms of  the vNE \cite{bus,bus2} and a
comparison with previous results lead to the conclusion that  the
LE can be an efficient and still valid entanglement measure for
binary  collisions,    as for other physical phenomena of interest
in quantum-information processing  where identical particles are
involved.

The paper is organized as follows. In Sec.~\ref{pri} we describe  the main properties of the LE as an
entanglement measure for two-fermion systems. In Sec.~\ref{pri2}
we evaluate numerically the time evolution of the LE
for a scattering event between a free propagating and a bound
electron in a two-dimensional system considering different spin
configurations. Conclusions are drawn  in
Sec.~\ref{pri3}.

\section{The theoretical criterion and its evaluation in a 2$N$-modes system}\label{pri}
In this section we  introduce the  entanglement criterion  for two-fermion systems
 based on the concept of   LE, usually applied to distinguishable particles.
Furthermore we  compare quantitatively  such a criterion   with
the one based on the vNE  in the case of  a simple system with
$2N$ degrees of freedom.

A pure state of two fermions can be written as \cite{Sch,schli2}:
\begin{equation}\label{pisso}
|\Psi_{F}\rangle=\sum_{i,j}^{2N}\omega_{ij}a^{\dag}_{i}a^{\dag}_{j}| 0\rangle
\end{equation}
where  $a_{i}$ and  $ a^{\dag}_{i}$ are the annihilation and  creation
operators of the mode $i$ satisfying  the usual fermionic
(anti)commutation  rules $\{a_{i},a^{\dag}_{j}\}=\delta_{ij}$,
and  $| 0\rangle $    is the vacuum state.  $\omega_{ij}$ are the
elements of  a complex and antisymmetric ($2N\,\times\,2N$) matrix
$\Omega$ where    $2N$  is the total number of modes for each
single particle,  while the normalization condition is given by
$\textrm{Tr}[\Omega^{\dag}\Omega]=1$. The single-particle reduced
density matrix $\rho$ for  the state $|\Psi_{F}\rangle$ can be
computed from the two-particle density matrix
$\rho_{F}=|\Psi_{F}\rangle\langle\Psi_{F}|$ and its elements are
\cite{You}:
\begin{equation}\label{fiu}
\rho_{\mu \nu}=\frac{\textrm{Tr}[\rho_{F}a_{\nu}^{\dag}a_{\mu}]}{
\textrm{Tr}[\rho_{F}{\sum_\mu}a_{\mu}^{\dag}a_{\mu}]}=(\Omega^{\dag}\Omega)_{\nu \mu}
\end{equation}
The  eigenvalues  of $\rho$  are  $|z_i|^{2}$, while the coefficients $z_i$  stem from the Schmidt decomposition of $|\Psi_{F}\rangle$
in terms of  Slater determinants \cite{Sch}. Furthermore it should be noticed that the eigenvalues of the one-particle reduced density matrix are pairwise identical and therefore it holds
$|z_{2k}|^{2}=|z_{2k-1}|^{2}$ with  $ \quad 1\leq k \leq N  $.
The number of coefficients $z_k$  that are  different from zero is the so-called
Slater rank, which can be related to the entanglement as follows:
a state with Slater rank equal to 1 (i.e. that  can be written as
a single Slater determinant) is  non-entangled, a state with Slater rank greater than 1 is  a linear combination of two or more Slater
determinants,  therefore  it can be considered  entangled.

Many alternative ways of defining a function apt at evaluating the lack of knowledge about a subsystem have been
proposed in the literature  among them the Tsallis entropy \cite{Tsallis} generalizes the concept of the vNE, encompassing,
among the others LE. It is defined as \cite{Tsallis2} :
\begin{equation} \label{tsal}
\varepsilon_{\scriptscriptstyle{q}}=\frac{1}{q-1} \textrm{Tr}\{\rho -\rho^{q}\}
\end{equation}
where $q$ is a real, not necessary positive, number. In the case of $q$ tending to 1 one obtains
the well known vNE  $\varepsilon_{\scriptscriptstyle{vN}}= -\textrm{Tr}\{\rho \ln{\rho}\}$ which satisfies
some standard properties as concativity, additivity  and sub-additivity, and which is acknowledged to be  a good quantum correlation measure 
 of a pure two-fermion state. When  $q$ is equal to 2   Eq.~(\ref{tsal}) reduces to 
$\varepsilon_{\scriptscriptstyle{L}}=1-\textrm{Tr}\rho^{2}$, that is  the LE  \cite{man}. 
In this paper  we  focus  on such a quantity  as a   measure of the  entanglement in systems  of 
indistinguishable particles. Even if  LE  is not additive in the usual sense as shown in Ref.\cite{man}, it has   some interesting
properties so far  not fully exploited. In fact it turns out to be extremely valuable  for the application of numerical methods,   as detailed in the following.

In terms of LE the quantum entanglement  of the pure two-fermion state  $|\Psi_{F}\rangle$ defined in Eq.~(\ref{pisso})
is given by 
\begin{equation} \label{kkkii}
\varepsilon_{\scriptscriptstyle{L}}=1-\sum_{i,j}^{2N}\left|\sum^{2N}_{l}\omega_{il}\omega_{lj}^{\ast}\right|^{2}.
\end{equation}
From    the above expression  we observe  that  the evaluation of
the LE can be performed   directly from the matrix $\Omega$ thus , in  numerical calculations, the definition and
allocation of the one-particle density matrix $\rho$ are  not
required: therefore $\varepsilon_{\scriptscriptstyle{L}}$    is
much easier to calculate than the vNE since no diagonalization of
the matrix  $\rho$ is needed. This aspect appears to be relevant
since the complexity of many systems of physical interest practically prevents the diagonalization of the corresponding density matrix.
Furthermore we note  that   Eq.~(\ref{kkkii})  is represention-independent since   the trace is invariant with respect to
unitary single-particle transformations.

By using  the above trace operations we can also express
$\varepsilon_{\scriptscriptstyle{L}}$ in terms of the eigenvalues
$|z_i|^{2}$ of the one-particle density matrix $\rho$  in order to compare
the expression of the LE  with the one of  the  vNE. By taking into account the above mentioned property for $|z_i|^{2}$, we find for the former that
\begin{equation} \label{bebel}
\varepsilon_{\scriptscriptstyle{L}}=1-\sum_{k}^{N}2|z_k|^{4}
\end{equation}
while, as shown in the literature \cite{You,ghira,sand}, the latter  can be written as
\begin{equation} \label{bebel2}
\varepsilon_{\scriptscriptstyle{vN}}= \ln{2}-\sum_{k}^{N}2|z_k|^{2} \ln{2|z_k|^{2}}.
\end{equation}
We stress that  Eq.~(\ref{bebel}) has been reported in oder to make more explicit the following discussion, but it is not employed
in the numerical calculations where  Eq.~(\ref{kkkii}) is used instead, not requiring the calculations of  the $\Omega$ eigenvalues.
In spite of similarities between the two expressions  for a two-fermion system, $\varepsilon_{\scriptscriptstyle{L}}$
 and  $\varepsilon_{\scriptscriptstyle{vN}}$  have a different dependence upon  the coefficients $z_k$.
Both Eqs.~(\ref{bebel}) and (\ref{bebel2})  attain their minimum
value when the state $|\Psi_{F}\rangle$ can be written  in terms
of a single Slater determinant. In this case we have
$|z_1|^{2}=\frac{1}{2}$  while  all the other coefficients are
zero and therefore
$\varepsilon_{\scriptscriptstyle{L}}=\frac{1}{2}$ and
$\varepsilon_{\scriptscriptstyle{vN}}=\ln{2}$. The fact that the
minimum value of the two measures is not  0, differently from what
happens for the distinguishable particles case, is  related to the
unavoidable correlations due to the exchange symmetry. Since the
quantum correlations related  only to antisymmetrization  of the
state of two fermions cannot be used to violate Bell's inequality
and are not a  resource for quantum-information processing (as
shown in previous works \cite{ghira,ghira2}) a  state with Slater
rank equal to 1  can be considered as non-entangled . As a
consequence we will assume that a value
$\varepsilon_{\scriptscriptstyle{L}}=\frac{1}{2}$  indicates a
non-entangled state.

For a maximum correlated state  it holds $|z_k|^{2}=\frac{1}{2N}$  $\forall\,\, k$   and $\varepsilon_{\scriptscriptstyle{L}}= 1-\frac{1}{2N}$.  We note that in the case
of a two-fermion system with a very large number $2N$ of modes, the maximum value of the LE  tends asymptotically  to 1 with  a power law as  for distinguishable particles.

In order to compare some properties of the LE   and vNE here we
shall analyze the two entanglement criteria for  a simple
2$N$-modes two fermion system in a  state $|\chi \rangle$, that
can be expressed  by  Eq.~(\ref{pisso}), with   the  following
coefficients of the antisymmetric matrix $\Omega$:
\begin{equation}\label{ficci}
\mathrm{\omega_{ij}}=\left \{ \begin{array}{lll}
  \sqrt{\frac{1+(N-1)(1-\alpha)^{2}}{2N}}  &\quad \textrm{for} \quad i=1,j=2    \\
 \sqrt{\frac{\alpha(2-\alpha)}{2N}} &\quad\textrm{for} \quad i=2k-1,j=2k  \quad\textrm{with}  \quad 2\leq k \leq N   \\
 0    &\quad \textrm{otherwise}
\end{array}\right .
\end{equation}
where $\alpha$ is a real parameter ranging between 0 and 1. In particular we observe that for $\alpha=0$
   the elements of the matrix  vanish  except for  $\omega_{1 2}$ which reduces to $\sqrt{\frac{1}{2}}$.
In this case  the state $ |\chi \rangle$ can be set  in terms of a
single Slater determinant and therefore  is  non-entangled. On the
other hand when $\alpha=1$ the condition of maximum entanglement
is reached, with all  the non vanishing coefficients equal
 to $\sqrt{\frac{1}{2N}}$.

\begin{figure}
    \includegraphics[width=0.6\linewidth]{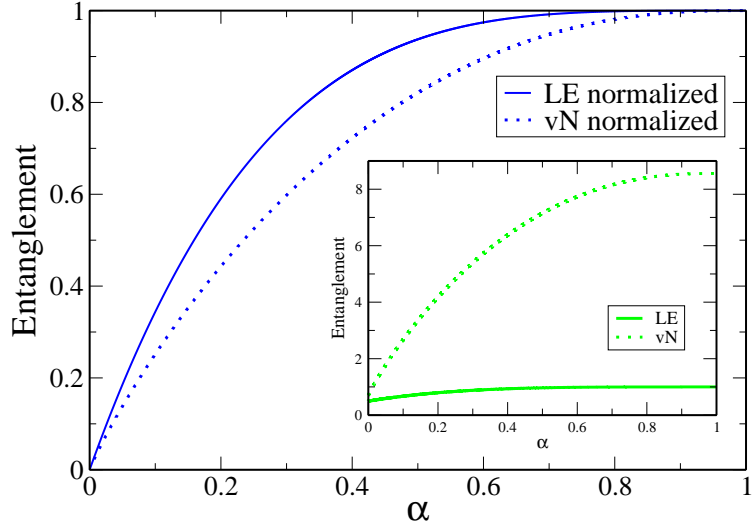}
   \caption{\label{fig_1} (Color online) Comparison between the normalized values of the  vNE and the LE as  a function   of the real parameter $\alpha$ in the 2$N$ modes two-fermion system  
described in the text (for this calculation
$N$ has been taken equal to 2601). The inset displays the dependence of the two not normalized measures  upon  $\alpha$. }
\end{figure}

In the above model the one-particle reduced density matrix $\rho$ is simply a
diagonal matrix with eigenvalues $ \omega_{12}^{2}$ and $
\omega_{2k-1,2k}^{2}$. This makes straightforward  the calculation
of the matrix  $\rho^{2}$. Its trace, needed for  the evaluation
of the LE ,  can be written as function of the
parameter $\alpha$ as:
\begin{equation}
\varepsilon_{\scriptscriptstyle{L}}^{ \chi }= 1-\frac{1}{2N}-\frac{(1-\alpha)^{4}(N-1)}{2N}.
\end{equation}
As expected for  $\alpha=0$,  $\varepsilon_{\scriptscriptstyle{L}}^{\chi}$ takes its minimum value $\frac {1}{2}$, while the maximum  value of $1-\frac {1}{2N}$  is reached for $\alpha=1$.
The expression of  $\varepsilon_{\scriptscriptstyle{L}}$   normalized to 1 reads
\begin{equation}
\tilde{\varepsilon}_{L}^{ \chi }=1-(1-\alpha)^{4} .
\end{equation}
The vNE  can also be  easily calculated and its
normalized form is
\begin{eqnarray}
\lefteqn{\tilde{\varepsilon}_{\scriptscriptstyle{vN}}^{\chi }= \frac{-1}{N \ln{N}} \bigg( (N-1)\alpha(2-\alpha) \ln{\frac{\alpha(2-\alpha)}{N}}{}} \nonumber \\
+{} & &\left(1+(1-\alpha)^{2}(N-1)\right)\ln{\frac{1+(1-\alpha)^{2} (N-1)}{N}} \bigg).
\end{eqnarray}
We note  that the  normalization of the vNE and of the LE allows
us to
 compare quantitatively the two measures  of the quantum entanglement. In  Fig.~\ref{fig_1}
we report the   entanglement of the system as a function of the
real parameter  $\alpha$. Both  curves    get
their  minimum value 0  for $\alpha=0$,  both  increase  with
$\alpha$ and reach 1 when $\alpha$ gets to 1. From the comparison
we note  that the LE  is always greater than the vNE as a
consequence of the different normalization procedures
apart from the initial and final values when they coincide. For
$\alpha$ tending to 1 (i.e. for $|\chi\rangle$ tending to the
maximally entangled state) the two measures attain  the same value
as  for the case of distinguishable particles \cite{Tzu}. For the sake of completeness we show the two non-normalized curves in the inset
of Fig.~\ref{fig_1}.

\section{Linear entropy dynamics in a two-electron scattering}\label{pri2}
\subsection{The model}
The dynamics of quantum entanglement has been investigated  in
various physical phenomena including, for example,     ionization
processes \cite{Feredov}, binary collision events
\cite{Jia,bus,Tal} and phonon-atom interaction \cite{chan}. Due to
the increasing quest for quantum computing capable solid state
devices, the study of the entanglement formation in a   scattering
event in semiconductor structures plays  an important role. A
numerical analysis of the entanglement dynamics in terms of vNE  for a two-electron collision in two-dimensional
semiconductor nanostructures   (GaAs) has been recently presented
in literature \cite{bus,bus2} analyzing the evolution of the
entanglement when an electron freely propagating interacts through
a screened Coulomb potential with another electron bound to a
specific site by a harmonic potential. In this section we intend
to study the entanglement   for such a  model  by using the
 criterion based on the LE  described in the previous section.

\begin{figure}
    \includegraphics[width=0.6\linewidth]{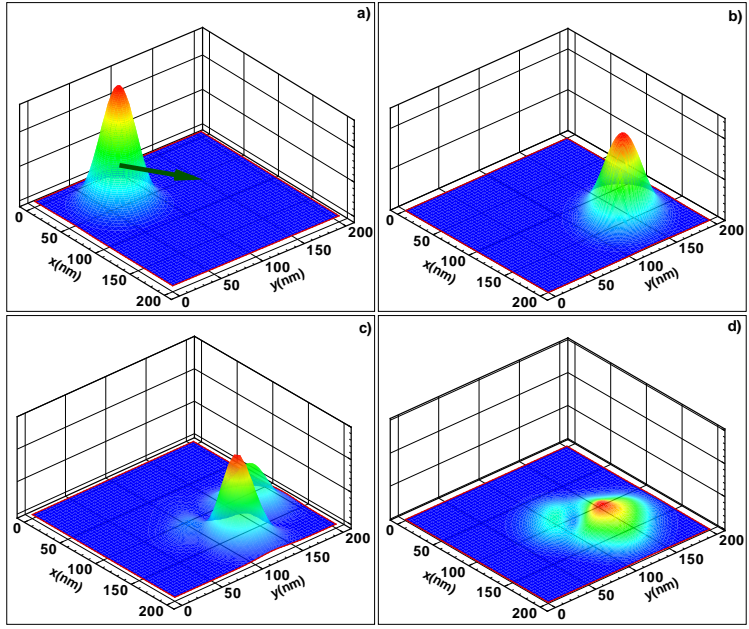}
   \caption{\label{fig_2} (Color online) Square modulus of the projection of the antisymmetrized two-particle wavefunction
 on the first three energy eigenstates $\xi_{n}$
of the harmonic oscillator centered in $\textbf{r}_{0}$=(95 nm, 95 nm): $\gamma_n(\textbf{r})=| \int d\textbf{r}'\xi_{n}(\textbf{r}')\Psi (\textbf{r},\textbf{r}')|^2$.  The  two upper graphs show
the projection on the ground state at two different times $t=0$ (a) and $t=480$ fs (b), while in the lower graphs the projections  on first (c) and the second (d) excited state
at time $t=480$ fs are reported.}
\end{figure}

The Hamiltonian describing the physical system can be written as:
 \begin{eqnarray}\label{rest2}
H(\textbf{r}_{a},\textbf{r}_{b})= - \frac{\hbar^{2}}{2m}\left(\frac{\partial^{2}}{\partial \textbf{r}_{a}^{2}}+
\frac{\partial^{2}}{\partial \textbf{r}_{b}^{2}}\right)+\frac{e^{2}}{\epsilon |\textbf{r}_{a}-\textbf{r}_{b}|}
+\frac{1}{2}m\omega^{2}(\textbf{r}_{a}-\textbf{r}_{0})^{2}  +\frac{1}{2}m\omega^{2}(\textbf{r}_{b}-\textbf{r}_{0})^{2}
\end{eqnarray}
where $\epsilon$ and $m$ are  the GaAs dielectric constant and
effective mass, respectively, and $\textbf{r}_{0}$ is the  center
of the harmonic potential, with  energy-level spacing
$\hbar\omega$. Spin-orbit effects have not been considered. At the
initial time  $t_{0}$ one of the two particles, namely the
incoming electron,  is represented by a minimum uncertainty
wave-packet centered in $\textbf{r}_{0}$(see Fig.~\ref{fig_2}a):
\begin{equation} \label{rosi}
\psi(\textbf{r},t_{0})= \frac{1}{\sqrt{2 \pi}\sigma}\exp{\left(
-\frac{(\textbf{r}-\textbf{r}_{0})^{2}}{4 \sigma^{2}}+i \textbf{k}\cdot \textbf{r}\right)}
\end{equation}
 where $\sigma$ is the mean spatial dispersion, $k=\sqrt{2 m
E_{k}}/\hbar $ with  $m$  the effective mass of the carrier and
$E_{k}$ is the carrier kinetic energy.
The bound electron is in the ground state of a  two-dimensional  harmonic oscillator
\begin{equation} \label{bisc1}
\phi(\textbf{r},t_{0})= \left(\frac{m\omega}{\pi \hbar}\right)^{1/2}\exp{\left( -\frac{m\omega(\textbf{r}-\textbf{r}_{1})^{2}}{2\hbar}\right)}
\end{equation}
where $\textbf{r}_{1}$ is the center of the harmonic potential, with energy-level spacing $\hbar\omega$.
The distance $|\textbf{r}_{1}-\textbf{r}_{0}|$ is such that   at  the initial time $t_{0}$  the Coulomb energy
 is negligible. We stress that, as a   consequence of the scattering,
 the state of the particle confined  in the harmonic oscillator (HO)  changes and that the two-particle
 wavefunction can be expressed as a single Slater determinant
 $\Psi(\textbf{r},\textbf{r}')=\psi( \textbf{r})\,\phi(\textbf{r}')    - \phi(\textbf{r}) \,\psi(\textbf{r}')$,
 only at the initial time. In order to better analyze
 the  state of the bound particle,   Figs.~\ref{fig_2}b-d display    the square modulus of the projection  of the  antisymmetrized two-particle wavefunction
$\Psi(\textbf{r},\textbf{r}')$  at time $t=480$ fs on the first
three eigenstates $\xi_{n}(n=1,2,3)$ of the harmonic oscillator:
$\gamma_n(\textbf{r})=| \int d\textbf{r}'\xi_{n}(\textbf{r}')\Psi
(\textbf{r},\textbf{r}')|^2$. Note  that the spectral
decomposition    on the  HO ground state at  time $t=0$
(Fig.~\ref{fig_2}a),   representing   the square modulus of the
one-particle initial wavefunction of the free carrier,  has been
almost totally transmitted  with no reflected part at $t=480$ fs .
The scattering event leaves the HO in a superposition of excited
states, as can been seen in Figs.~\ref{fig_2}c and \ref{fig_2}d
and for   higher energies the peaks of the  function
$\gamma_n(\textbf{r})$ are closer to the center of the HO. This is
due to the different energies  of the outgoing particle as the
bound particle is left in different HO excited states. Note that
the Coulomb potential also creates spatial correlations between
the two electrons.

 We  consider  now the effect of different initial  spin configurations on the evolution of the entanglement.
 In the first quantum state studied  the two electrons have the same spin (spin up):
\begin{equation}\label{res}
|\Psi\rangle=\frac{1}{\sqrt{2}}\bigg(|\psi \,\phi\rangle   - |\phi \,\psi\rangle \bigg)|\!\!\uparrow \uparrow \rangle
\end{equation}
where    the  wavefunctions corresponding to the states $|\psi\rangle$ and $|\phi\rangle$ are of the type defined in Eqs.~(\ref{rosi}) and (\ref{bisc1}), respectively, and
  the ket  $|\!\! \uparrow \rangle$ indicates a  spin up state. The  explicit form
of  the matrix $\Omega$ for the state  $|\Psi\rangle$ can be
obtained by discretizing the spatial coordinate $\textbf{r}$ into
a $N^{2}$ points grid ($N$ points for each spatial dimension) as
shown in Ref.~\cite{bus}. We get for the  $2N^{2} \times 2N^{2}$
matrix $\Omega_{\psi}$
\begin{equation}
\mathrm{\Omega_{\Psi}}=\frac{1}{\sqrt{2}}\left( \begin{array}{cc}
\Omega_{A} & 0\\
0  &  0 \\
\end{array}\right),
\end{equation}
where $\Omega_{A}$ is the antisymmetric $N^{2} \times N^{2}$  matrix
whose elements  read
\begin{equation}
\omega_{ij}=\psi( \textbf{r}_{i})\,\phi(\textbf{r}_{j})    - \phi(\textbf{r}_{i}) \,\psi(\textbf{r}_{j}).
\end{equation}

 The second case  considered is the one
with two electrons having different spins, that  cannot be
factorized  in a spin and a real  space      term. The form of the
two-particle state is
\begin{equation} \label{drog}
   |\Upsilon \rangle=\frac{1}{\sqrt{2}}    \bigg(|\psi\, \phi\rangle    |\!\! \uparrow  \downarrow \rangle  - |\phi\, \psi\rangle
|\!\! \downarrow \uparrow \rangle\bigg).
\end{equation}
By applying  a   procedure analogous to the one used for $|\Psi\rangle$ and  introducing an unitary transformation for the spin variables \cite{bus}, we get
\begin{equation}
\mathrm{\Omega_{\Upsilon}}=\frac{1}{2}\left( \begin{array}{cc}
\Omega_{A} & -\Omega_{S}\\
\Omega_{S}  &  -\Omega_{A} \\
\end{array}\right).
\end{equation}
where $\Omega_{S}$ is the symmetric counterpart of  $\Omega_{A}$.

The last  two  states  considered still describe   two electrons with different spins, but contrary to the previous case they
 can be factorized in a position   term and in a spin  term. We can identify  the singlet spin state
\begin{eqnarray}\label{quest}
|\Phi\rangle=\frac{1}{2}\bigg(|\psi \,\phi\rangle    + |\phi \,\psi\rangle \bigg)
\bigg(| \!\!\uparrow \downarrow  \rangle -|   \!\!\downarrow \uparrow \rangle\bigg)
\end{eqnarray}
with
\begin{equation}
\mathrm{\Omega_{\Phi}}=\frac{1}{2}\left( \begin{array}{cc}
0 & -\Omega_{S}\\
\Omega_{S}  & 0  \\
\end{array}\right),
\end{equation}
and the triplet  spin state
\begin{equation}\label{quest1}
|\Xi\rangle=\frac{1}{2}\bigg(|\psi \,\phi\rangle    - |\phi \,\psi\rangle \bigg)
\bigg(|\!\!\uparrow \downarrow  \rangle +|\!\!\downarrow \uparrow \rangle\bigg)
\end{equation}
with
\begin{equation}
\mathrm{\Omega_{\Xi}}=\frac{1}{2}\left( \begin{array}{cc}
\Omega_{A} & 0\\
0  &  -\Omega_{A} \\
\end{array}\right).
\end{equation}
The LE  of the triplet state can be easily obtained
from the one of the same-spin state $|\Psi\rangle$
\cite{ghira,bus}.  In fact for the eigenvalues of the one-particle
reduced density matrix
 of the state   $  |\Xi \rangle$ it holds
\begin{equation}
  |z_{i}^{ \Xi }|^{2}=|z_{i+N^2}^{ \Xi }|^{2}=\frac{1}{2}|z_{i}^{ \Psi }|^{2}\quad\textrm{for}  \quad 1\leq i \leq N^2 .
\end{equation}
Therefore from the Eq.~(\ref{bebel}) its  LE is given by
\begin{equation} \label{bebel3}
\varepsilon_{L}^{ \Xi }=1-2\sum_{i}^{N^2}|\frac{z_i^{ \Psi
}}{2}|^{4}=\frac{1}{2} (1+\varepsilon_{L}^{ \Psi }).
\end{equation}
The property expressed by the relation~(\ref{bebel3})  is due  to
the fact that the triplet state, like the singlet state, cannot be
factorized in a space term and a spin term and  this property
remains true  during the time evolution. From  Eq.~(\ref{bebel3})
we observe that the minimum value  of the LE  of     $
|\Xi \rangle$ is $\frac{3}{4}$, greater than the value of
$\frac{1}{2}$ corresponding to a non-entanglement condition for
any two-fermion state according the criterion introduced in the
previous section. Such an offset  is related  to   the lack of
knowledge about the spin of the particles. The  behavior shown by
the triplet  and singlet states  is in agreement with the one
obtained  in previous works where  the entanglement formation is
evaluated  in terms of the   vNE  for  the case of
one-dimensional and  two-dimensional scattering  \cite{bus,bus2},
as we show in the following.

\subsection{Numerical results}
In order to calculate  the entanglement dynamics in the   system
described above, we solve numerically  the time-dependent
Schr\"{o}ndiger equation for the two-particle wavefunction
considering  as initial condition two electrons described by the wave-packets given in  the Eqs.~(\ref{rosi}) and  (\ref{bisc1}). In this way at
each time step we have  the two-particle  wavefunction needed to
define  the matrix  $\Omega$. Finally  from
expression~(\ref{kkkii}) and  by using only the matrix elements
$\omega_{ij}$ we can    compute the time evolution of the
entanglement in terms of the LE . We stress that such
an approach is  simple and  does not require matrix
diagonalization procedures which result to be very demanding from
the point of view of the numerical calculation.

\begin{figure}
    \includegraphics[width=0.6\linewidth]{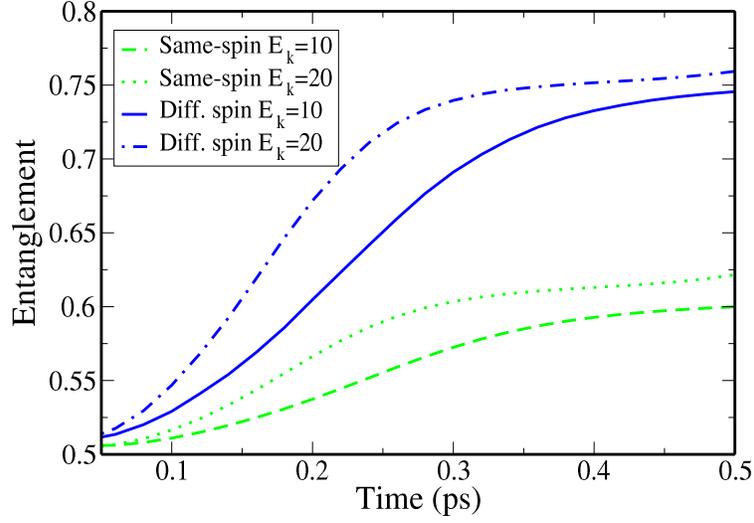}
   \caption{\label{fig_3} (Color online) Entanglement as a function of the time for different initial  states:  same-spin state $|\Psi\rangle$
for two different initial energy values of the  incoming electron
$E_k =$ 10 meV (dashed  line) and  20 meV (dotted line); the
state with  electron having  different spin $ |\Upsilon \rangle$
for $E_k =$ 10 meV (dot-dashed line) and 20 meV (solid  line). The
harmonic oscillator energy is $\hbar \omega =$ 2 meV.}
\end{figure}

Figure~\ref{fig_3} shows that at initial time the LE
for the states $|\Psi\rangle$ and $|\Upsilon \rangle$ get its
minimum value $\frac{1}{2}$. In fact the Coulomb energy is still
negligible being the two wave-packets far enough,  and the only
quantum correlations present are due to the exchange symmetry.
Therefore, as expected,  $|\Psi\rangle$ and $|\Upsilon \rangle$
must be considered as initially non entangled. We observe that as
the free carrier get closer to the center of the harmonic
potential the quantum correlation builds up and the LE 
reaches a stationary value once the particles get far enough. Such
a value depends upon  the initial kinetic energy of the
propagating  carrier  $E_k$: in particular  it is higher for
higher energies  for both spin configurations, in good qualitative
agreement with the previous results found for scattering events
between two distinguishable particles \cite{Bordone,Bertoni, Gun}.

\begin{figure}
    \includegraphics[width=0.6\linewidth]{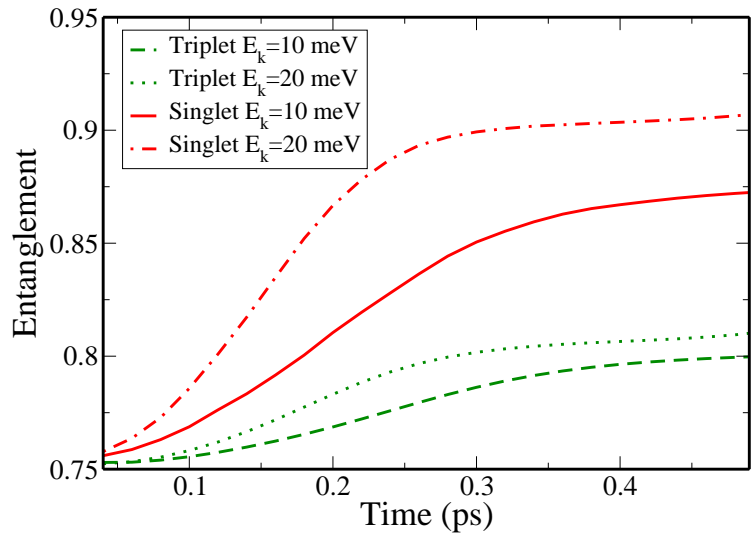}
   \caption{\label{fig_4} (Color online) Entanglement as a function of the time:  triplet spin state $|\Xi\rangle$ for two different initial energy values of the
 incoming electron  $E_k =$ 10 meV (dashed  line) and  20 meV (dotted line)
and  singlet spin state $|\Phi\rangle$ for  $E_k =$ 10 meV (dot-dashed line) and 20 meV (solid  line). The harmonic oscillator energy is $\hbar \omega =$ 2 meV.}
\end{figure}

In Fig.~\ref{fig_4}  the time evolution of the entanglement for
the singlet $|\Phi\rangle$ and triplet $|\Xi\rangle$ spin states
are presented. As expected,  at  the initial time  the LE is $\frac{3}{4}$. This implies that $|\Phi\rangle$ and
$|\Xi\rangle$ are initially entangled,  being  their Slater rank
greater than 1. In fact  from Eqs.~(\ref{quest}) and
(\ref{quest1})  we observe that they cannot be put in terms of a
single Slater determinant. Such a result confirms the
correspondence between the LE  and the Slater rank
criterion.

\begin{figure}
    \includegraphics[width=0.6\linewidth]{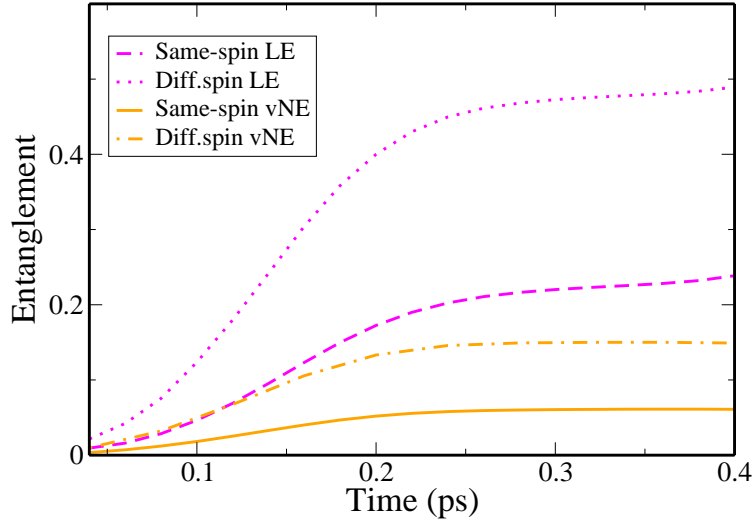}
   \caption{\label{fig_5} (Color online) Two measures  of the  entanglement as a function of the time.
     The time evolution of the normalized  LE  for same spin state $|\Psi\rangle$
 (dashed  line) and   for the state  with electrons having  different spin $ |\Upsilon \rangle$ (dotted line),
  is compared with  the time evolution of the normalized vNE for the same  states:  $|\Psi\rangle$
 (solid  line) and  $ |\Upsilon \rangle$ (dash-dotted line). The initial kinetic energy of the incoming electron is $30$ meV while the harmonic oscillator energy is $\hbar \omega =$ 2 meV.}
\end{figure}

In order to better compare     the   properties of the
entanglement measure for  an electron-electron scattering
obtained using the
 LE and vNE,   the time evolution
of the entanglement, evaluated according to the two criteria,  is
presented in  Fig.~\ref{fig_5} for the states  $|\Psi\rangle$ and
$|\Upsilon \rangle$. As  for the case of the theoretical model
studied in the previous section, here  we have normalized the two
measures in order to compare them quantitatively. At initial time
both of them    are zero
  since   no quantum correlation   is  initially present apart from the one related to the exchange symmetry. Then, at increasing times,
  for a given state
the LE  is always greater than the vNE.
Nevertheless the time of the entanglement formation, defined as
the time at which the entanglement reaches its  stationary value,
is the same for  both   measures. This  behavior can be ascribed  to
the fact that, as indicated by the  Eqs.~(\ref{bebel}) and
(\ref{bebel2}), both the  LE and vNE   can be expressed as
function of $|z_k|^2$, the eigenvalues of the one-particle reduced
density matrix, which  can be assumed  weakly time-dependent for
large times.

\section{Conclusion}\label{pri3}
In the last years the  notion  of the entanglement for  systems of
identical particles has been widely discussed: various approaches
have been proposed in literature    each having advantages and
drawbacks  \cite{Zanardi,Sch,Fisher,You,Vacca}. In this paper we
have analyzed  one possible  theoretical criterion to quantify
 the quantum correlations appearing in a two-fermion systems. It can be considered the generalization
of the LE, usually adopted in the context of
distinguishable particles. Such a criterion is based on the
fermionic analogous of the Schmidt decomposition theorem and uses
the one-particle reduced density matrix. In particular our
analysis shows that LE  permits to determine
whether the uncertainty concerning the states derives only  from
the  indistinguishability of the particles or it is  a  genuine
manifestation of the entanglement. This aspect is crucial since, as shown in  previous works, the quantum  correlation due
to the exchange symmetry does  not represent a good resource for
quantum information processing\cite{ghira,ghira2}.

 The criterion  proposed here appears to be  closely related to the one involving
the evaluation of the vNE  of the reduced
statistical operator \cite{You,ghira,sand}. To compare the two
criteria and analyze their properties  we have quantified the
entanglement according to the two measures
 in a simple theoretical model describing a two-fermion system with 2$N$ degrees of freedom. We found that the
LE is greater than  the vNE, but  the two measures give the  same
value when the quantum state of the system tends  to the maximally
entangled one   in agreement with results obtained for
distinguishable particles \cite{Tzu}.

Furthermore we have  used the LE  to quantify the  time
evolution of quantum correlations   in a numerically simulated
electron-electron collision  event. In agreement with the previous
analyses obtained by using the vNE  \cite{leon,bus},
our numerical results  show that the entanglement dynamics depends
on the spin components of the states even if  the Hamiltonian does
not include spin-terms. Moreover also in this case the triplet and
singlet spin states are initially  entangled and such an
entanglement can be ascribed to the lack on knowledge about the
spin state of a particle in a specific real-space state. For our
two-electron model  the values of the entanglement obtained  by
using the LE results to be higher than the ones obtained by using
the vNE in agreement  with what found analytically for a simple model of a
$2N$-mode two-fermion  system. Most notably we found that
the time of the entanglement formation is the same for the two
measures.

Finally    we  note that    the calculation of the entanglement in
terms of the LE  is  easier  and computationally much faster than the one performed by means of the
 vNE since no  diagonalization of the one-particle reduced density matrix is   required.  Therefore   we
believe  that the LE  is an useful  correlation measure for a two-fermion system and its application
turns out to be very helpful to investigate the entanglement dynamics   for    those  physical systems
 with a very large number of degrees of freedom,  whose complexity does not allow  the diagonalization
of the reduced statistical operators through  numerical procedures.

\begin{acknowledgments}
The authors would like to  thank Carlo Jacoboni  for useful discussions.
We  acknowledge support from the U.S Office of Naval Research (contract No. N00014-03-1-0289/ N00014-98-1-0777).
\end{acknowledgments}


\begin{thebibliography}{99}

\bibitem{Schr}
E.~Schrodinger,
  Naturwissenschaften  \textbf{23}, 807 (1935).




\bibitem{Giuli}
D.~Giulini et al., \emph{Decoherence and the Appearance of a Classical World in Quantum theory}
(Springer, Berlin, 1996.)




\bibitem{Peres}
A.~Peres,  \emph{Quantum Theory: Concepts and Methods}
(Kluwer Academy Publishers, The Netherlands, 1995.)


\bibitem{Sch}
J.~Schliemann, J.I.~Cirac, M.~Kus, M.~Lewenstein and D.~Loss,
 Phys. Rev. A \textbf{64}, 022303 (2001).


\bibitem{Shi}
Yu~Shi,
 Phys. Rev. A, \textbf{67}, 024301 (2003).

\bibitem{Li}
Y.S.~Li, B.~Zeng, X.S.~Liu and G.L.~Long,
 Phys. Rev. A, \textbf{64}, 054302  (2001).


\bibitem{Zanardi}
  P.~Zanardi,
 Phys. Rev. A, \textbf{65}, 042101(R)(2002).







\bibitem{Fisher}
J.R.~Gittings and A.J.~Fisher,
 Phys. Rev. A,  \textbf{66}, 032305 (2002).


\bibitem{You}
R.~Paskauskas and L.~You,
 Phys. Rev. A,  \textbf{64}, 042310 (2001).


\bibitem{Vacca}
H.M.~Wiseman and J.A.~Vaccaro,
 Phys. Rev. Lett.,  \textbf{91}, 097902  (2003).

\bibitem{Levay}
P.~Levay, S.~Nagy and J.~Pipek,
Phys. Rev. A, \textbf{72}, 022302 (2005).


\bibitem{brand}
Fernando G.S.L.~Brandao,
 Phys. Rev. A,  \textbf{72}, 022310  (2005).



\bibitem{zeng}
B.~Zeng, H.~Zhai and Z.~Xu,
 Phys. Rev. A,  \textbf{66}, 042324  (2002).


 \bibitem{dow}
M.R.~Dowling, A.C.~Doherty and H.M.~Wiseman,
 Phys. Rev. A,  \textbf{73}, 052323  (2006).

\bibitem{bus}
F.~Buscemi, P.~Bordone and A.~Bertoni,
 Phys. Rev. A,  \textbf{73}, 052312  (2006).





\bibitem{leon}
L.~Lamata and J.Leon,
 Phys. Rev. A,  \textbf{73}, 052322  (2006).

\bibitem{raa}
A.~Ramsak, I.~Sega and J.H.~Jefferson,
 Phys. Rev. A,  \textbf{74}, 010304(R)  (2006).




\bibitem{inn}
T.~Inn, C.~Ellenberger, K.~Ensslin, C.~Yannouleas, U.~Landman, D.C.~Driscoll, and A.C.~Gossard, 
e-print cond-mat/0610029.




\bibitem{schli2}
J.~Schliemann, D.~Loss, and A.H.~MacDonald,
 Phys. Rev. B \textbf{63}, 085311 (2001).


\bibitem{Schl3}
K.~Eckert, J.~Schliemann, D.~Bruss and  M.~Lewenstein,
Annals of Physics \textbf{299}, 88 (2002).


\bibitem{ghira}
G.C.~Ghirardi and L.~Marinatto,
 Phys. Rev. A  \textbf{70}, 012109 (2004).

\bibitem{sand}
X.~Wang and B.~Sanders,
  J. Phys. A: Math. Gen. \textbf{38}, 67 (2005).

\bibitem{ghira2}
G.~Ghirardi, L.~Marinatto and T.~Webber,
 J. Stat. Phys \textbf{108}, 49 (2002)
\bibitem{bus2}
F.~Buscemi, P.~Bordone and A.~Bertoni,
 Optics and Spectroscopy (to appear).

\bibitem{Tsallis}
C.Tsallis,
J. Stat. Phys. \textbf{52}, 479 (1988).

\bibitem{Tsallis2}
A.K~Rajagopal and R.W.~Rendell,
Europhysics news  \textbf{36/6}, 221 (2005).

\bibitem{man}
G.~Manfredi and M.R.~Feix
 Phys. Rev. E  \textbf{62}, 4665  (2000).

\bibitem{Tzu}
Tzu-Chieh Wei  \emph{et  al.},
Phys. Rev. A  \textbf{67}, 022110 (2003).

\bibitem{Feredov}
M.V.~Fedorov \emph{et  al.},
Phys. Rev. A  \textbf{69}, 052117 (2004).

\bibitem{Jia}
Jia Wang, C.K.~Law and M-C~Chu,
Phys. Rev. A  \textbf{73}, 034302 (2006).
\bibitem{Tal}
A.~Tal  and G.~Kurizki,
 Phys. Rev. Lett.  \textbf{94}, 160503, (2005)




\bibitem{chan}
W.H.~Chan and C.K~Law,
 Phys. Rev. A \textbf{74}, 024301  (2006).




\bibitem{Bordone}
P.~Bordone and A.~Bertoni,
 J.Comp.Elec  \textbf{3}, 407 (2004).

\bibitem{Bertoni}
A.~Bertoni,
 J.Comp.Elec. \textbf{2}, 291 (2003).




\bibitem{Gun}
D.~Gunlycke, J.H.~Jefferson, T.~Rejec, A.~Ramsak, D.G.~Pettifor and  G.A.D.~Briggs,
J. Phys.: Condens. Matter \textbf{18},  S851-S866 (2006).







\end{thebibliography}
\end{document}